\documentclass[11pt,a4paper]{article}
\pdfoutput=1

\usepackage{bm}
\usepackage{amsmath,amssymb}
\usepackage{cite}

\usepackage{graphicx}
\usepackage{color}

\usepackage{hyperref} 
\hypersetup{colorlinks=true, citecolor=blue, filecolor=black, linkcolor=blue, urlcolor=blue, pdfpagemode=UseNone}

\numberwithin{figure}{section}
\numberwithin{equation}{section}

\newcommand{\be}{\begin{equation}}
\newcommand{\ee}{\end{equation}}
\newcommand{\bea}{\begin{eqnarray}}
\newcommand{\eea}{\end{eqnarray}}

\usepackage[normalem]{ulem}

\newcommand{\dir}{}

\newcommand{\phys}[1]{#1} 
\newcommand{\cR}{\bar{\mathcal{R}}}
\newcommand{\hR}{\hat{\mathcal{R}}}
\newcommand{\cK}{\mathcal{K}}
\newcommand{\Lap}{\bar{\Delta}}

\newcommand{\physical}{\mathrm{phys}}
\newcommand{\sR}{\mathfrak{R}}

\newcommand{\gmn}{g_{\mu\nu}}

\newcommand{\bgmn}{\bar g_{\mu\nu}}
\newcommand{\hmn}{h_{\mu\nu}}

\newcommand{\hess}{\bar{\Gamma}^{(2)}}

\newcommand\ie{\textit{i.e.}\ }
\newcommand\eg{\textit{e.g.}\ }
\newcommand\cf{\textit{cf.}\ }

\newcommand{\aka}{{a.k.a.}\ }

\newcommand{\viz}{{\it viz.}\ }
\newcommand{\half}{\tfrac{1}{2}}

\newcommand{\eps}{\varepsilon}

\begin{document}

\begin{titlepage}

\begin{center}
{\huge \bf Large curvature and background scale independence in single-metric approximations to asymptotic safety} 
\end{center}
\vskip1cm


\begin{center}
{\bf Tim R. Morris}
\end{center}

\begin{center}
{\it STAG Research Centre \& Department of Physics and Astronomy,\\  University of Southampton,
Highfield, Southampton, SO17 1BJ, U.K.}\\
\vspace*{0.3cm}
{\tt  T.R.Morris@soton.ac.uk}
\end{center}

\abstract{In single-metric approximations to the exact renormalization group (RG) for quantum gravity, it has been not been clear how to treat the large curvature domain beyond the point where the effective cutoff scale $k$ is less than the lowest eigenvalue of the appropriate modified Laplacian. 
We explain why this puzzle arises from background dependence, resulting in Wilsonian RG concepts being inapplicable.
We show that when properly formulated over an ensemble of backgrounds, 
the Wilsonian RG can be restored.  
This in turn implies that solutions should be smooth and well defined no matter how large the curvature is taken.
Even for the standard single-metric type approximation schemes, 
this construction can be rigorously  derived by imposing a modified Ward identity (mWI) corresponding to rescaling the background metric by a constant factor. However compatibility in this approximation requires the space-time dimension to be six. Solving the mWI and flow equation  simultaneously, new variables are then derived that are independent of  overall background scale.}

\end{titlepage}

\tableofcontents

\newpage

\section{Introduction}
\label{sec:Intro}

When applied to quantum gravity, asymptotic safety is the idea that  the Wilsonian renormalization group (RG) flow of gravitational couplings approaches a viable interacting non-perturbative fixed point in the far ultraviolet, such that physical observables are rendered ultraviolet finite despite perturbative non-renormalisability \cite{Weinberg:1980}.
Ever since a functional (\aka ``exact'' \cite{Wilson:1973})
RG equation adapted to this case, was put forward in ref. \cite{Reuter:1996}, a steady increase of interest in the asymptotic safety programme for quantum gravity has produced a wealth of results. For reviews and introductions see \cite{Reuter:2012,Percacci:2011fr,Niedermaier:2006wt,Nagy:2012ef,Litim:2011cp}. 

However,  in order to actually calculate anything, some approximations have to be made. A frequent approximation is to retain only a finite number of local operators in the effective action. These `polynomial truncations' can therefore can be viewed as built on a small curvature expansion. Here `small' means with respect to the effective cutoff scale $k$. If we write such terms in dimensionless form using $k$, then such a truncation is really only justified  if these terms remain much less than one. 
For example writing dimensionless (\aka scaled) scalar curvature as $\bar{\sR} = \phys{\bar{R}}/k^2$, we require $\bar{\sR}\ll1$. (For further discussion on this point, see ref. \cite{DietzMorris:2013-1}. $\bar{R}$ is the physical curvature, while a bar indicates that the metric $\bgmn$ is the background metric.)

In order to go beyond this in a substantive way, it is necessary in effect to keep an infinite number of such local operators. Then it is possible to treat them without expansion, at least within some model approximations, and thus explore properties which are invisible to polynomial truncations, for example singularities at finite scaled curvature, or 
 scaling laws or asymptotic behaviour
when the scaled curvature is diverging. Asymptotic safety if it makes sense, must also make sense in these regimes.  

Here $k$ plays the r\^ole of an infrared  (IR) cutoff imposed by hand on the eigenvalues $\lambda$ of some appropriate modified Laplacian for fluctuation fields $u$:
\be 
\Lap\, u = \lambda^2 u\,.
\ee
Schematically,  $\Lap = -\bar{\nabla}^2 + \bar{E}$, where $\bar{E}$ is some endomorphism depending on the background metric.
%
We will initially assume that the cutoff is sharp, although we will shortly address the general case. The problem we address arises when 
the minimum eigenvalue is positive \cite{Demmel:2014fk,Demmel2015b,Ohta2016,Falls:2016msz}. It will prove helpful in this case, as shown, to parametrise them as $\lambda^2$, and refer to $\lambda$ (taken positive) as the eigenvalue. Then we have the properties that there is a minimum eigenvalue $\lambda>0$, and $k$ and $\lambda$ have the same mass dimension so can be directly compared.

Let us now note that for a Wilsonian RG `step' to be well defined, it must be possible to lower $k$ to any strictly positive value, 
without encountering singularities. 
The true partition function is only recovered when the limit $k\to0$ is taken, removing the cutoff.
The following apparent paradox then arises in the case of interest. What meaning do we attach to a large curvature regime where $k$ is smaller than any eigenvalue? On the one hand the passage $k\to0$ at fixed physical curvature, corresponds to exploring ever larger dimensionless curvature, and we have already noted that the solutions must continue to be smooth for any positive $k$ if the RG is to remain well defined \cite{DietzMorris:2013-1}. On the other hand, once $k$ is less than any eigenvalue there is nothing left to cut off and thus imposing smoothness criteria on solutions at arbitrarily large dimensionless curvature would appear to be physically meaningless \cite{Demmel:2014fk,Demmel2015b,Ohta2016}.\footnote{We mean smooth in the precise sense of continuously infinitely differentiable, for example with respect to $\bar{\sR}$.}

\begin{figure}[ht]
\centering
\includegraphics[scale=0.25]{\dir{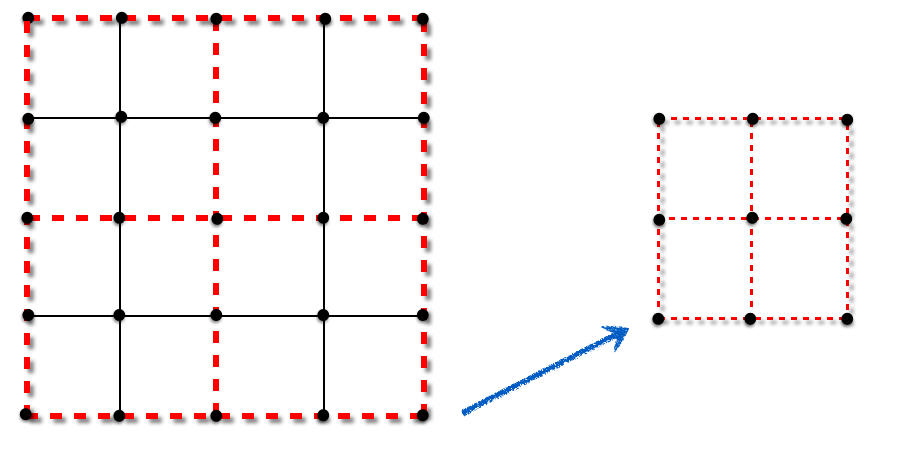}}\hskip1cm
\includegraphics[scale=0.3]{\dir 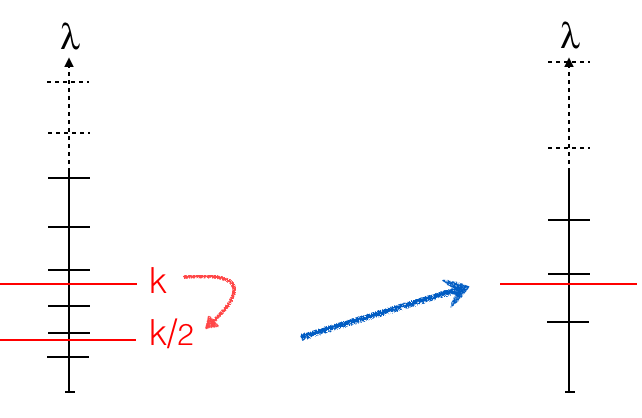}
\caption{A $4\times4$ lattice is blocked to twice the lattice spacing. After rescaling back to the original lattice spacing, it cannot be the same lattice, since it is now only $2\times2$. Likewise, a system with a lowest eigenvalue $\lambda$ cannot be the same after integrating out from $k$ to $k/2$. After rescaling $\lambda$ so that the IR cutoff again has value $k$, there are less eigenvalues below $k$ than when we started.}
\label{fig:finiteSystem}
\end{figure}

At first sight this appears to be just a technical conundrum, albeit without any clear resolution.
Actually, we can view it as a fundamental impasse which should never have been encountered in a meaningful application of the Wilsonian RG. To see this, let us go back to basics. A Wilsonian RG transformation consists of two steps: a Kadanoff blocking transformation \cite{Kadanoff:1966wm}, for example of a lattice of spins to a courser one of twice the lattice spacing, followed by a rescaling of dimensions to bring the system back to its original size \cite{Wilson:1971bg,Wilson:1973,WegnerBook1}. Universal behaviour flows from fixed points. And fixed points require that after rescaling, the basic lattice structure itself looks exactly the same. For example we cannot meaningfully formulate the Wilsonian RG for a strictly finite lattice (see fig. \ref{fig:finiteSystem}). After Kadanoff blocking, the lattice has less cells,
so no rescaling will make it look exactly the same.\footnote{Let us note in passing that in the limit of large lattices, the deviation from universality can be quantified in certain ``finite size effects'', although such technology will not be relevant here.}

The reader will see in fig. \ref{fig:finiteSystem} that we have an analogous issue. After a Kadanoff blocking, \eg $k\mapsto k/2$,  we can go from a situation where there were eigenvalues remaining to be cutoff, to, for example, one where there are no eigenvalues remaining to be cutoff. No rescaling will make these situations look the same. Wilsonian RG concepts such as fixed points are thus not applicable. This is particularly glaring around the lowest eigenvalue, but of course it is not there that from this point of view the RG ceased to have any real meaning. It never really made sense at any finite $k$. For example on a compact space with a discrete set of eigenvalues, for any finite $k=k_1$ there is some finite number $N_1$ of eigenvalues remaining to be integrated out.\footnote{We remind the reader that the IR cutoff $k$ is equivalent to the effective UV cutoff of a Wilsonian effective action \cite{Morris:1993,Morris:2015oca} and thus to integrating out modes above $k$.}  On lowering $k$ to $k=k_2$, this number reduces to $N_2<N_1$, so again it impossible to rescale the blocked system to make it the same as the one at $k_1$.

One cannot escape this problem by working on a non-compact space such as a Euclidean hyperboloid. In this case the spectrum is not discrete but there is still a lowest eigenvalue that sets a scale and there is also an integrable density of eigenvalues $\rho(\lambda)$ \cite{Camporesi:1994ga}, see also \cite{Falls:2016msz,Benedetti:2014gja}. The analogous situation then arises in that the dimensionless integral of $\rho(\lambda)$ over the remaining range of eigenvalues is reduced. One also cannot escape this problem by using a smooth cutoff profile that only suppresses modes with lower eigenvalues, rather than sharply cutting off the fluctuations. After integrating out from $k_1$ to $k_2$, less of the lower modes are suppressed to the same extent, reflecting the new position for $k$, and again no rescaling of dimensions can untie this.

Therefore from this perspective, the Wilsonian RG itself 
cannot meaningfully be formulated on such a space for any curvature or any value of $k$. While the results we report apply to any such situation, we are particularly interested in 
the so-called $f(\sR)$ approximation \cite{Machado:2007,Codello:2008,Benedetti:2012,DietzMorris:2013-1,DietzMorris:2013-2,Falls:2013bv,Demmel:2014hla,Demmel:2014fk,Falls:2014tra,Demmel2015b,Eichhorn:2015bna,Ohta:2015efa,Ohta2016,Falls:2016msz}, where all powers of the  scalar curvature are kept and summarised in a Lagrangian of form $f_k(\bar{\sR})$. 
In the literature this is in fact so far the only example where the functional form is treated without expansion.
In order to understand at the intuitive level what has gone wrong, let us recall that the curvature is that of a fixed metric, the background metric $\bar{g}_{\mu\nu}$, chosen to be a maximally symmetric Euclidean space (we will take the typical choice of a sphere and thus positive curvature) and the eigenvalues for fluctuations of the metric, $h_{\mu\nu}$, are those of the (modified) Laplacian formed from this background metric.\footnote{The utilisation of the single-metric approximation (addressed below) obscures the difference between $\bgmn$ and the total metric $\gmn$. The important point here is that it is a fixed metric that we choose to input.} 
The picture on the right in fig. \ref{fig:finiteSystem}, is the one we see when considering linearised fluctuations about this fixed $\bar{g}_{\mu\nu}$. The scale of the eigenvalues $\lambda$ is set by the background curvature $\phys{\bar{R}}$.

\begin{figure}[ht]
\centering
\includegraphics[scale=0.35]{\dir 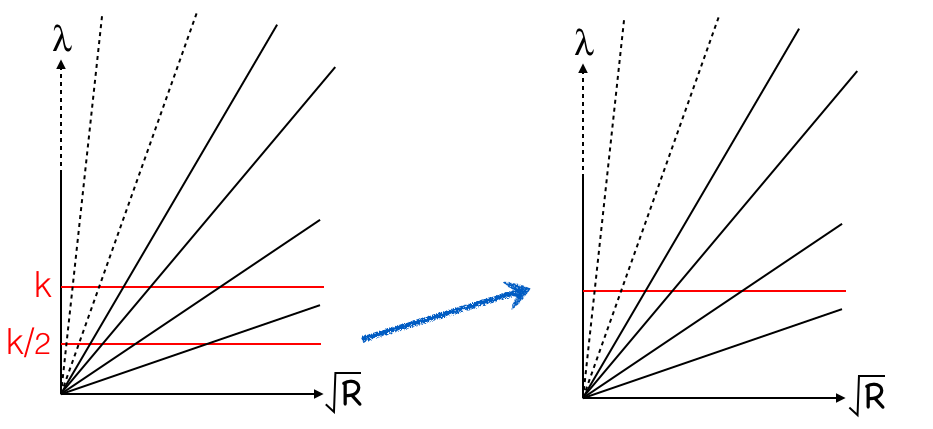}
\caption{A continuous ensemble of spheres has eigenvalues $\lambda\propto \phys{\sqrt{R}}$.  Blocking by integrating out from $k$ to $k/2$, and then rescaling $\lambda\mapsto 2\lambda$ such that the IR cutoff is again at $k$, also rescales the horizontal axis: $\phys{\sqrt{R}}\mapsto 2\phys{\sqrt{R}}$. The graph therefore remains invariant, and the distribution of eigenvalues below $k$ is unchanged by this RG step.}
\label{fig:allR}
\end{figure}

As already anticipated in the conclusions of ref. \cite{Dietz:2016gzg}, the problem arises because  \emph{background independence} is not respected. By  \emph{background independence}, we mean that physics should not depend on the choice of background, but instead depend only on the full metric $g_{\mu\nu}$  \cite{Dietz:2015owa}. 
(Typically a linear split is considered so that $g_{\mu\nu}=\bar{g}_{\mu\nu}+ h_{\mu\nu}$, although there are exceptions, \eg \cite{Ohta:2015efa,Ohta2016}. Our arguments here are independent of how the split is performed.) Since in the partition function, the full metric (directly or through
$\hmn$)  is integrated over, a continuous infinity of manifolds is actually included. The eigenvalues in fig. \ref{fig:finiteSystem}, correspond to just one of these. Even if we choose to restrict to Euclidean spheres, we should still be integrating over their size. By ranging over this ensemble, the scalar curvature $\phys{R}$ thus takes all positive values.  The Laplacian formed using the full metric, will thus yield eigenvalues that, in the ensemble, form a continuum, since they themselves depend on the curvature. From the perspective of the full metric the situation can be illustrated as in fig. \ref{fig:allR}. As we see from the figure, the basic precondition for the Wilsonian RG is then restored, namely that the structure of the full ensemble of eigenvalues can look exactly the same after an RG step. 

To see in detail how this ensemble repairs the problem, write the lowest eigenvalue as $\lambda = a \phys{\sqrt{R}}$, where $a$ is a pure number. The set $\mathcal{S}_k$ of spheres with $\phys{R}>k^2/a^2$ have no eigenmodes left to integrate out. Let 
us focus on a sphere $s$ with one particular physical curvature $\phys{R}=R_s$. It is true that in the region $a\sqrt{R_s}< k < 2a\sqrt{R_s}$, a blocking $k\mapsto k/2$ will take us from a situation where this sphere had fluctuations to integrate out, to one where it no longer has fluctuations to integrate out. However there is now also a sphere with curvature $\phys{R}=R_s/4$ that after blocking and rescaling, looks exactly the same as the original sphere $s$. Meanwhile under blocking, the original sphere $s$, together with all spheres in the range $k^2/4a^2<\phys{R}< k^2/a^2$, join the now enlarged set $\mathcal{S}_{k/2}$ of spheres with no modes left to integrate out. Despite the addition of new members, this set is isomorphic to the original, since under the rescaling of all mass dimensions by two, it turns back into $\mathcal{S}_k$. (Indeed the set can be described in a $k$-independent way by indexing the spheres by their scaled curvature $\sR>1/a^2$ rather than their physical curvature $\phys{R}$.)

In order to adapt the infrared cutoff employed in constructing the flow equation \cite{Nicoll1977,Wetterich:1992,Morris:1993}, and to gauge fix, the background field method is employed and this is why the full metric $g_{\mu\nu}$ and background metric $\bar{g}_{\mu\nu}$ are introduced \cite{Reuter:1996}. 
We have just seen that the confusion over the r\^ole of large dimensionless curvature $\sR$, and in particular whether constraints on the solution apply to all $k$ or only $k>a\sqrt{R}$, is resolved by properly incorporating \emph{background independence}. In the literature the construction of the effective action about a general background metric $\bgmn$, and thus computing in effect on all backgrounds simultaneously, is also referred to as background independence.\footnote{As explained in ref. \cite{Reuter:2008qx}, this usage follows that in loop quantum gravity \cite{Ashtekar:1991hf,Ashtekar:2004eh,Rovelli:2004tv,Thiemann:2007zz}.} If we regard the Wilsonian RG applied to the ensemble above as effectively a Wilsonian RG applied simultaneously to a continuous ensemble of spheres with different \emph{background} metrics --related by overall scale, then indeed again we see we have a resolution to the conundrum. We see that it makes no sense to give preferential treatment to a region of $k<a\sqrt{\bar{R}}$, making $k$ thus dependent on the background metric. The Wilsonian RG framework is correctly recovered only if all background metrics, whatever their overall scale, are treated democratically, \ie with the process of integrating-out being functionally independent of the value of $\bar{R}$. We thus conclude that a solution must remain smooth for all $k>0$, and thus we must impose that solutions $f(\sR)$ remain smooth no matter how large $\sR$ is taken.

We are forced to treat the ensemble of spheres in this way because it is required by the full functional integral, once we recognise that treating the fluctuations $\hmn$ around background metric $\bgmn$ is equivalent to treating spheres with different full metrics $\gmn$. This equivalence is enforced by \emph{background independence} in the sense we mean it \cite{Bridle:2013sra,Dietz:2015owa}, where in fact it is a strong \emph{extra} constraint. We know in principle how to recover this requirement through imposition of modified split Ward identities \cite{Pawlowski:2005xe,Litim:2002hj,Bridle:2013sra,Reuter:1997gx,Litim:1998nf,Litim:2002ce,Manrique:2009uh,Manrique:2010mq,Manrique:2010am,Becker:2014qya,Dietz:2015owa,Labus:2016lkh}. 
In full gravity it is a challenging task to satisfy the full modified split Ward identity, facing not just practical problems but potentially problems of principle \cite{Labus:2016lkh}. In practice it is broken by single metric approximations of the type that give us the $f(\sR)$ approximations we have just been considering. However at the intuitive level at which we have so far been operating, we can ignore this and regard the problem of large curvature as solved.

In the rest of the paper we will show that nevertheless it is possible to make the arguments more rigorous, even using the standard approximations, although as a consequence of the approximations this will only be achieved in $d=6$ space-time dimensions.
The intuitive argument 
points the way. A weaker version of background independence, one we will
 call \emph{background rescaling invariance}, is sufficient for our purposes. We need to use the appropriate modified Ward identity (mWI) to restore the link between one mode of the linearised fluctuation, namely $h_{\mu\nu}\propto \bgmn$, and rescaling the metric and thus the size of the background sphere. Once these ideas have been fully developed, we will be able to show that, essentially, the $f(\sR)$ approximations already constructed in the literature \cite{Machado:2007,Codello:2008,Benedetti:2012,DietzMorris:2013-1,DietzMorris:2013-2,Falls:2013bv,Demmel:2014hla,Demmel:2014fk,Falls:2014tra,Demmel2015b,Eichhorn:2015bna,Ohta:2015efa,Ohta2016,Falls:2016msz}, can be reinterpreted in a background independent way.

Although we have in mind addressing such approximations, in particular projecting on a maximally symmetric background and using the optimised cutoff \cite{opt1,opt3}, we prove this for any background metric $\bgmn$ describing a compact space-time, and for any choice of cutoff profiles $r_u$ (that is including different profiles for different fluctuation fields if desired). 

At first sight the prospects for such a reinterpretation seem remote. Firstly, the $f(\sR)$ truncation crucially relies on the single-metric approximation, which amounts to identifying $\gmn$ and $\bgmn$ at an appropriate point in the calculation. In contrast, we must use the full bi-metric approximation which retains both fields. \textit{A priori} there is no reason to expect approximations in this approach to look anything like the single-metric results. We will see how we are able to make contact with these within an appropriate approximation.
 Secondly we must face head on the problem that gauge fixing itself breaks background independence \cite{Reuter:1996}, which means it cannot be recovered in general without 
taking $k\to0$ and going on-shell \cite{Dietz:2015owa,Labus:2016lkh}. Fortunately 
the mWI we are aiming for, one related to metric rescaling, in effect only rescales the value of the gauge fixing parameter in this sector. We will see in sec. \ref{sec:GF}
that since Landau gauge is taken (setting the parameter to zero), after negotiating some subtleties, this change drops out. Thirdly, generically in any uncontrolled approximation, the mWI will prove to be incompatible with the flow, leading to an overconstrained system with no solutions \cite{Labus:2016lkh}. Fortunately our mWI 
is sufficiently simple to escape this danger, but only if we choose $d=6$. 

In the next section, we develop background rescaling invariance as an exact symmetry, deriving and solving the corresponding Ward identity, in particular treating the fluctuation $\hmn$ also through its York decomposition \cite{York:1973ia,Dou:1997fg,Lauscher:2001ya,Codello:2007bd}, and introducing the average physical scalar mode $\bar{h}$ which will play a crucial r\^ole in the arguments that follow. 

Even when we break the symmetry by gauge fixing and adding cutoffs, we are able to make progress by keeping the analysis at a high level, \ie without specifying the form of the background metric $\bgmn$, or the detailed form of the approximation. In this way, we will see that the arguments take on a  particularly clean and elegant form. 
In sec. \ref{sec:symmetries} we see one aspect of this, where we show how to compute the effects of background rescaling by trading it for diffeomorphism invariance and dimensional analysis. 

In sec. \ref{sec:GF}, we introduce gauge fixing. It is possible to recover background rescaling invariance if we take the Landau gauge limit, as is commonly done. We will also handle the determinants that arise from the change of variables to the York decomposition. We will see that the key to extending background rescaling invariance to these sectors is assigning appropriate background-scaling dimensions, \aka indices $d_u$, to various fluctuation fields. 

In sec. \ref{sec:IR} we introduce the IR cutoffs in the standard way considered in the literature  \cite{Demmel:2014fk,Demmel2015b,Ohta2016,Falls:2016msz}. Since they depend only on the background metric, they turn out to have simple scaling behaviour under background rescaling, apart from a correction that takes into account that the IR cutoff scale $k$ is invariant under this. 

This gives us all we need to derive the modified Ward identity (mWI) in sec. \ref{sec:mWI}. We see in the final equation of this section the first intimations of why $d=6$ dimensions is special for background rescaling invariance. In sec. \ref{sec:compatibility} we explain why this is required when uncontrolled expansions are considered, in particular why we must have compatibility of the approximated mWI with the approximated flow equation. 

In sec. \ref{sec:single-metric} we define a suitable single-metric type approximation which however retains dependence on $\bar{h}$. In sec. \ref{sec:compatibility-singlemetric} we prove that within this approximation the mWI and flow equation are compatible with each other if and only if $d=6$, independent of the choice of cutoff profiles and $\bgmn$. In sec. \ref{sec:six} we pause for a moment to give an intuitive explanation of the significance of $d=6$ dimensions, in particular we see that if we had based the theory on a four-derivative action (such as for Weyl gravity), $d=8$ dimensions would be singled out. 

Then in sec. \ref{sec:background-independence} we see that, in common with previous cases  \cite{Bridle:2013sra,Dietz:2015owa,Labus:2016lkh}, the approximate flow equation and mWI can be solved simultaneously to reveal some hidden variables, $\hat{g}_{\mu\nu}$ and $\hat{k}$, 
which in this case are independent of the overall scale of the background metric. We then show that in terms of these variables, precisely the single-metric approximations in the literature are recovered. A crucial element in this and the proof in sec. \ref{sec:compatibility-singlemetric}, is the proof that in precisely $d=6$ dimensions, the natural action for the Hessians can be shown order by order to be independent of $\bar{h}$. 

Finally in sec. \ref{sec:conclusions}, we see that such variables do indeed describe the ensemble solution to regaining the Wilsonian RG, as we sketched above.

\section{Background rescaling invariance, Ward identity and solution}
\label{sec:WI}

We start by deriving the unbroken Ward identity and then solving it by the method of characteristics. We operate at a formal level for now, \ie we will not worry about gauge fixing, regularisation, and the effect of approximations. These will be added in secs. \ref{sec:GF}, \ref{sec:IR}, and \ref{sec:single-metric} respectively.  
Our discussion in this section might seem overly expansive at points but the reader will later see how the observations made here, and the equations derived here, become key to understanding background rescaling invariance when all the above complications are folded in.

We begin by expanding the full quantum metric $\gmn$ in terms of the background metric $\bgmn$ and fluctuation field $\hmn$ as
\be 
\label{total}
\gmn = \bgmn + \hmn\,,
\ee
so that the partition function takes the form:
\be 
\label{Z}
\mathcal{Z}[\bgmn,J^{\alpha\beta}]=\int \!\!\mathcal{D}\hmn \, 
\exp\left\{-S_0[\bgmn\!+\!\hmn]+\int \!J^{\mu\nu}\hmn\right\}\,.
\ee
(Here $S_0$ is the bare action. We work in $d$ dimensional space-time with Euclidean signature. By $\int$ on its own we mean $\int\!d^dx$. We have absorbed the $\sqrt{\bar{g}}$ factor into our definition of the source $J^{\mu\nu}$ which is thus a tensor density of weight -1. The reader may prefer to keep the $\sqrt{\bar{g}}$ factor explicit at the expense of some extra terms at intermediate stages. Since $J$ disappears in the end, the end result is the same.)

We are interested in implementing \emph{background rescaling invariance}, \ie making explicit the fact that under a rescaling of the background metric, 
\be
\bgmn \mapsto (1-2\eps)\, \bgmn\,, \label{rescalebg-1} 
\ee
the total metric \eqref{total}, and thus also the physics, does not change, if at the same time we compensate by changing the fluctuation field as follows:
\be 
\label{rescale1-1}
\hmn \mapsto \hmn +2\eps\,\bgmn\,.
\ee
For our purposes we need only the case where $\eps$ is space-time independent, and furthermore we take it infinitesimal. (The factor $2$ is immaterial but will prove convenient later.) Since each term in the background Levi-Civita connection $\bar{\Gamma}^\mu_{\alpha\beta}$ contains one background metric and its inverse, $\bar{\Gamma}^\mu_{\alpha\beta}$ is then invariant under \eqref{rescalebg-1}. Thus the background Riemann and Ricci curvatures are also invariant, while the background scalar curvature transforms to
\be 
\bar{R}\mapsto (1+2\eps) \,\bar{R}\,.
\ee
Choosing a space of constant scalar curvature (typically a Euclidean sphere), the transformations \eqref{rescalebg-1} and \eqref{rescale1-1}  thus have the desired effect of making explicit that physics should not depend on the value of this background curvature.

From here on we clean up the notation and write background rescaling invariance more simply as 
\be
\delta\bgmn = -2\, \bgmn\,, \label{rescalebg} 
\ee
where it is to be understood that the RHS (right hand side) is multiplied by an arbitrary constant infinitesimal proportionality factor ($\eps$) which then drops out of the final formulae. Similarly we write 
\be 
 \label{rescale1}
\delta \hmn =+2\,\bgmn
\ee
and
\be 
\label{rescaleR}
\delta \bar{R}=2\,\bar{R}\,.
\ee
Writing $W = \ln\mathcal{Z}$, we thus have that a change of background in form \eqref{rescalebg}, compensated by a change of integration variable as in \eqref{rescale1}, leads only to a shifted source term in \eqref{Z} and thus
\be 
-\int\! \bgmn\, \frac{\delta W}{\delta \bgmn} = \int\! J^\mu_\mu
\ee
(where the index on $J$ is lowered using the background metric). Introducing the Legendre effective action $\Gamma[\bgmn,\hmn]$ via
\be 
W = -\Gamma + \int\! J^{\mu\nu}\hmn\,,
\ee
where $\delta \Gamma/\delta \hmn = J^{\mu\nu}$ and  $\hmn$ now refers to the classical field 
$\hmn = {\delta W}/{\delta J^{\mu\nu}}$, we thus find the Ward identity
\be 
\int \bgmn \left( \frac{\delta\Gamma}{\delta\hmn}-\frac{\delta\Gamma}{\delta\bgmn}\right)=0\,.
\ee
This equation can be solved by the method of characteristics. Thus from 
\be 
\int \left(\delta\hmn \frac{\delta\Gamma}{\delta\hmn}+ \delta\bgmn \frac{\delta\Gamma}{\delta\bgmn}\right) -\delta\Gamma=0\,,
\ee
we identify the normal to the solution surface to be the vector
\be 
\left[\frac{\delta\Gamma}{\delta\hmn}, \frac{\delta\Gamma}{\delta\bgmn},-1\right]\,,
\ee
and thus the vector field that generates characteristic curves depending on some auxiliary parameter $t$ to be (again we keep a factor 2 for later convenience):
\be 
\label{charGamma}
\frac{\partial}{\partial t}\Gamma = 0\,,
\ee
\be 
\label{charbg}
\frac{\partial}{\partial t} \bgmn(x,t) = -2\, \bgmn(x,t)
\ee
and
\be 
\label{charh1}
\frac{\partial}{\partial t} \hmn(x,t) = 2\,\bgmn(x,t).
\ee
These equations are easily solved to obtain that $\Gamma$ is $t$-independent,
\be 
\label{solbg}
\bgmn(x,t) = {\rm e}^{-2t}\,\bgmn(x,0)
\ee 
and
\be 
\hmn(x,t) = \gmn(x) -\bgmn(x,t)\,.
\ee
In the above solution we have recognised that the $x$ dependent integration constant can appropriately be called the classical total metric. Thus (see also ref. \cite{Labus:2016lkh} and the appendix to ref.
\cite{Dietz:2015owa}) we deduce that $\Gamma$ must only be a functional of the $t$-independent combination 
\be 
\label{gmn-invariant}
\gmn(x) = \bgmn(x,t) + \hmn(x,t)\,.
\ee
Of course we knew this all along, but it is encouraging to see that such a global background rescaling invariance (\ref{rescalebg},\ref{rescale1}) alone is already sufficient to enforce this.

In reality we are not interested in working directly with $\hmn$, but following common practice we want to make a York (\aka transverse traceless) decomposition \cite{York:1973ia,Dou:1997fg,Lauscher:2001ya,Codello:2007bd}:
\be \label{TT}
h_{\mu\nu} = h_{\mu\nu}^{T} + {\bar\nabla}_\mu \xi_\nu + {\bar\nabla}_\nu \xi_\mu + {\bar\nabla}_\mu {\bar\nabla}_\nu \sigma + \frac{1}{d} { \bar g}_{\mu\nu} h \, ,
\ee
where $\xi_\mu$ and $\sigma$ are the gauge degrees of freedom to be distinguished from the physical traceless-transverse mode $h^T_{\mu\nu}$ and physical scalar mode $h$, $\bar{\nabla}_\alpha$ is the background-covariant derivative, and
\be
\label{TTdefs}
h_{\mu}^{T\,\mu} = 0 \, , \quad {\bar\nabla}^\mu h_{\mu\nu}^{T} = 0
\, , \quad {\bar\nabla}^\mu \xi_\mu = 0 \, , \quad { h} = h^\mu_\mu -{\bar\nabla}^2 \sigma \,.
\ee
While such a decomposition is important in the computations we will study, a price to pay is that \eqref{gmn-invariant} will no longer be the only background rescaling invariant  combination.
We see that in terms of the York decomposition, the fluctuation transformation \eqref{rescale1} becomes
\be 
\label{rescale2}
\delta h = 2\,(h + d)
\ee
We already see that the other terms in \eqref{TT} are invariant, clearly so for $h^T_{\mu\nu}$ but also for the gauge degrees of freedom\footnote{This will change however as a result of the gauge fixing in sec. \ref{sec:GF}.} since $\bar{\nabla}_\alpha$ is invariant. 
Either by change of variables using \eqref{TT}, or repeating the initial analysis, we find that the Ward identity is replaced by:
\be 
\label{hWI}
\int  \left( (d+h)\frac{\delta\Gamma}{\delta h}-\bgmn\frac{\delta\Gamma}{\delta\bgmn}\right)=0\,,
\ee
and thus, solving this, \eqref{charGamma} and \eqref{charbg} remain the same but \eqref{charh1} is replaced by
\be 
\frac{\partial}{\partial t} h(x,t) = 2d +2h(x,t)\,.
\ee
This equation has general solution
\be 
\label{solh}
h(x,t) = -d +\left( h(x,0) + d\right) {\rm e}^{2t}\,.
\ee
Using \eqref{solbg}, this implies that
\be 
\label{sol2}
\left(1+\frac{h(x,t)}{d}\right)\bgmn(x,t) = \left(1+\frac{h(x,0)}{d}\right)\bgmn(x,0)
\ee
is an invariant. Adding in the other manifest invariants from \eqref{TT}, we see again that $\gmn(x)$ is also invariant. Thus we see that, as a result of the York decomposition, we now have several invariant variables, namely $h^T_{\mu\nu}, \xi_\mu, \sigma$ and the combination \eqref{sol2}. 

When we choose the background to be of finite volume (a sphere for example), $h(x)$ has a normalisable constant mode $\bar{h}$, the zero mode of the Laplacian $-\bar{\nabla}^2$. We can decompose $h$ as
\be 
\label{h-expand}
h(x) = \bar{h}+h^\perp(x)\,,
\ee
where $h^\perp$ is orthogonal to $\bar{h}$, \ie $\bar{h}\int\!\sqrt{\bar{g}}\,h^\perp =0$.
The characteristic \eqref{solh} then also decomposes as $h^\perp(x,s) = h^\perp(x,0)\, {\rm e}^{2t}$, that scales multiplicatively, while $\bar{h}$ keeps the non-homogeneous pieces in \eqref{solh}:
\be 
\label{solhbar}
\bar{h}(t) = -d +\left( \bar{h}(0) + d\right) {\rm e}^{2t}\,,
\ee
or equivalently
\be 
\label{charhbar}
\frac{\partial}{\partial t} \bar{h}(t) = 2d +2\bar{h}(t)\,.
\ee
In terms of this decomposition, we therefore have that
\be 
\label{rescale3}
\delta h^\perp =2\,h^\perp\qquad{\rm and}\qquad \delta\bar{h}=2\, (\bar{h} +d)\,.
\ee
And we have the invariant combinations
$ \bgmn \,h^\perp$ and
\be 
\label{invariant}
\left(1+\frac{\bar{h}}{d}\right) \bgmn\,.
\ee
It is this last combination that will prove most useful. Its background rescaling invariance is most transparent if we recognise that from \eqref{rescale3}, 
\be 
\label{rescale4}
\delta \left(1+\frac{\bar{h}}{d}\right) = 2\left(1+\frac{\bar{h}}{d}\right) 
\ee 
transforms homogeneously. 

Taking the Einstein-Hilbert action as an example, the background rescaling invariant version is arrived at by replacing $\bgmn$ with \eqref{invariant} in $\sqrt{\bar{g}} \bar{R}$, giving
\begin{multline} 
\label{invariant-example}
\sqrt{\bar{g}} \bar{R} \left(1+\frac{\bar{h}}{d}\right)^{d/2-1} =
\sqrt{\bar{g}} \bar{R} + 
\frac{d-2}{2d}\sqrt{\bar{g}} \bar{R} \bar{h} + 
\frac{(d-2)(d-4)}{8d^2}\sqrt{\bar{g}}\bar{R} \bar{h}^2 \\
+\frac{(d-2)(d-4)(d-6)}{48d^3}\sqrt{\bar{g}}\bar{R} \bar{h}^3
+ O(\bar{h}^4)\,.
\end{multline}
We see that for general $d$, background rescaling invariance, \eqref{rescalebg} and \eqref{rescale3}, requires infinitely many interactions. The $\bar{h}$ independent part scales as $\delta (\sqrt{\bar{g}} \bar{R}) = (2-d) \sqrt{\bar{g}} \bar{R}$ under \eqref{rescalebg}, but this is cancelled by the inhomogeneous part of the transformation of the $O(\bar{h})$ part.  Indeed using \eqref{rescalebg} and \eqref{rescale3}, the $O(\bar{h})$ part transforms as
\be 
\label{O(h-bar)}
\frac{d-2}{2d}\delta(\sqrt{\bar{g}} \bar{R} \bar{h}) = (d-2) \sqrt{\bar{g}} \bar{R} - \frac{(d-2)(d-4)}{2d}\sqrt{\bar{g}} \bar{R} \bar{h}\,.
\ee
Likewise, the homogeneous part of this transformation is cancelled by the inhomogeneous part coming from the $O(\bar{h}^2)$ term and so on. 

We remark that for $d$ a positive even integer, there are in fact only finitely many interactions. In $d=2$ dimensions, there are no $\bar{h}$ interactions. This reflects the fact that $\int\! \sqrt{\bar{g}} \bar{R}$ is then a topological quantity (the Euler characteristic). In $d=4$ dimensions the series stops at $O(\bar{h})$, reflecting the fact that the higher order interactions for the physical scalar mode always contain at least one derivative.
In $d=6$ dimensions, we see that all cubic and higher terms in $\bar{h}$ vanish. The significance of this observation will become clear later.

\section{Relation to diffeomorphism invariance and dimensions}
\label{sec:symmetries}

It will prove useful to notice that we can intertwine the metric rescaling \eqref{rescalebg} with two other symmetries which are actually preserved exactly, namely (background) diffeomorphism invariance and the rescaling symmetry corresponding to dimensional assignments.\footnote{The interrelation of these symmetries has been also been discussed in refs.\cite{DietzMorris:2013-2,Dietz:2015owa}.} Using diffeomorphism invariance, 
the rescaling \eqref{rescalebg} can be achieved, within some coordinate patch,  by rescaling the coordinates:
\be 
\label{rescalex}
\delta x^\mu=\, x^\mu
\ee
(together with a change in the argument of the field). 
Indeed, diffeomorphism invariance in this case induces the tensor transformation 
\be 
\label{rescaleTensor}
\delta T_{\alpha_1\cdots \alpha_q}^{\beta_1\cdots\beta_p} =(p-q)\,  T_{\alpha_1\cdots \alpha_q}^{\beta_1\cdots\beta_p}\,.
\ee
However we can untie \eqref{rescalex} (and also return the argument of the field back to $x^\mu$) by recognising that we also have a multiplicative symmetry in theory space as a statement of mass dimensions. The fact that all the equations must be dimensionally correct tells us that 
\be 
\label{rescaleMass}
\delta Q =[Q]\,Q
\ee
must also be an invariance of the flow equations and modified Ward identities, where $Q$ is any quantity, and $[Q]$ is its mass dimension. 

Thus for any quantity \emph{whose field dependence is restricted to that of the background metric}, rescaling the background metric while leaving the coordinates alone, as in \eqref{rescalebg}, is equivalent to applying the diffeomorphism \eqref{rescaleTensor} followed by the dimensional rescaling \eqref{rescaleMass}. 

For example, applied to a scalar quantity such as the background scalar curvature, only the latter transformation operates. This is why \eqref{rescaleR} is the same result we would obtain from using \eqref{rescaleMass} and recognising that $\bar{R}$ is dimension two. 

A less trivial example is furnished by the modified Laplacian operator $\Lap$. In general this takes the form of the appropriate tensor operator on the modes $u$ we are considering (for example the Lichnerowicz Laplacian in the case of symmetric tensor modes) plus further modifications as desired (the endomorphism piece \cite{Demmel:2014fk,Demmel2015b,Ohta2016,Falls:2016msz}). However since such an
 operator $\Lap$ is a map from the space of modes $u$ back into the same space, and is furthermore constructed using only the background metric field, these operators behave like scalars as far as this discussion is concerned. Indeed, if they carry indices, they carry an equal number $p=q$ of upper and lower indices. It follows that their transformation law under \eqref{rescalebg}, 
also merely reflects dimensional assignments, and thus:
\be 
\label{rescaleDelta}
\delta\Lap=2\,\Lap\,.
\ee

\section{Background rescaling with gauge fixing and auxiliary fields}
\label{sec:GF}

The exact type and number of fluctuation fields $u$, depends on the details of the implementation \cite{Benedetti:2012,Demmel:2014hla,Demmel:2014fk,Demmel2015b,Ohta2016,Falls:2016msz}, however generically these include  versions of the component fields in \eqref{TT},  together with ghosts, and with auxiliary fields that arise from the change of variables to \eqref{TT}, or ultimately a subset of all these.

We have established in sec. \ref{sec:WI}, the form of the background rescaling invariance, generated by \eqref{rescalebg} with \eqref{rescale1} --  or equivalently with \eqref{rescale2} or \eqref{rescale3}.  However we ignored the infrared cutoff terms, gauge fixing, and auxiliary fields. In this section we show how the framework generalises when the latter two are taken into account.


Since the  auxiliary fields arise from characterising the measure (Jacobians) for fluctuations around $\bgmn$, their action is bilinear and 
 transforms only as induced by $\bgmn$ itself. The transformation does not depend on $\hmn$ or the rest of the action. In practice gauge fixing is  chosen to depend on $\bgmn$ and $\hmn$ alone (rather than being dependent on the detailed form of the action for example) and to be linear in $\hmn$. (This is discussed in more detail below.) Therefore like the auxiliary fields, the ghost action is also bilinear and its transformation law 
depends only on $\bgmn$ itself, and not on $\hmn$ or the rest of the action. It follows that the discussion of sec. \ref{sec:symmetries} applies and the kernels in these actions  transform homogeneously. We can then define the transformation laws of the ghosts and auxiliary fields to cancel this and make these actions invariant. 


The reader can verify these statements on their own favourite implementation of the ghost and auxiliary sectors. To make these considerations concrete here, we consider as an example the ghost and auxiliary fields on a maximally symmetric background as described in ref. \cite{Benedetti:2012}.  The ghost action is written there as:
\begin{multline} \label{ghosts2}
S_{\rm gh} = \int\!\! \sqrt{\bar{g}}\, \Big\{  \bar{C}^{T\mu} \Big( \bar{\nabla}^2 
+\frac{\bar{R}}{d}\Big)^2 C_\mu^T +4\Big( \frac{d-1}{d}\Big)^2 \bar{c} \Big( \bar{\nabla}^2 +\frac{\bar{R}}{d-1}\Big)^2 \left(-\bar{\nabla}^2\right) c \\
    +  B^{T\, \mu} \Big( \bar{\nabla}^2 +\frac{\bar{R}}{d}\Big)^2 B_\mu^T 
    +4 \Big( \frac{d-1}{d}\Big)^2 b \Big( \bar{\nabla}^2 +\frac{\bar{R}}{d-1}\Big)^2 \left(-\bar{\nabla}^2\right) b \Big\}\, ,
    \end{multline}
where the $C_\mu^T$ and $c$ are complex Grassmann fields, while $B_\mu^T$ and $b$ are real fields, and the index $T$ denotes transverse vectors,\footnote{We have however rescaled the ghost fields to absorb  an overall factor of $Z_k/\alpha$. Compared to ref. \cite{Benedetti:2012}, the same should be done  for the gauge dependent component fields $\xi_\mu$ and $\sigma$. The gauge parameter $\alpha$ is discussed below.}
 while the action for auxiliary fields reads:
\begin{multline} \label{aux-gr}
S_{\rm aux-gr} =  \int\!\! \sqrt{\bar{g}}\, \Big\{  2 \bar{\chi}^{T\, \mu} \Big( -\bar{\nabla}^2 -\frac{\bar{R}}{d}\Big) \chi_\mu^T + \Big( \frac{d-1}{d}\Big) \bar{\chi} \Big( \bar{\nabla}^2 +\frac{\bar{R}}{d-1}\Big)\bar{\nabla}^2 \chi    \\
    + 2 \zeta^{T\mu} \Big( -\bar{\nabla}^2 -\frac{\bar{R}}{d}\Big) \zeta_\mu^T +\Big( \frac{d-1}{d}\Big) \zeta \Big( \bar{\nabla}^2 +\frac{\bar{R}}{d-1}\Big) \bar{\nabla}^2 \zeta  \Big\}\, ,
\end{multline}
where the $\chi_\mu^T$ and $\chi$ are complex Grassmann fields, while $\zeta_\mu^T$ and $\zeta$ are real fields.
Finally the Jacobian for the transverse decomposition of the ghost action is given by
\be
\label{aux-gh}
S_{\rm aux-gh} = \int\!\! \sqrt{\bar{g}} \, \phi\left(- \bar{\nabla}^2\right) \phi \, .
\ee
We can either apply  \eqref{rescalebg} directly or recognise that, by the discussion of sec. \ref{sec:symmetries}, the kernels transform according to their dimension together with a correction from \eqref{rescaleTensor} when indices are raised (for example on $\bar{C}^T_\mu$). Either way, we readily read off the transformation law for the auxiliary and ghost fields that leaves these actions invariant. Writing 
\be 
\label{rescaleu}
\delta u = \frac{d-d_u}{2} u\,
\ee
where the first factor takes care of the volume term,
we see that for this implementation all the ghosts in \eqref{ghosts2} have $d_{\rm ghost}=6$, all the auxiliary fields in \eqref{aux-gr} have $d_{\rm aux-gr}=4$ and the ghost auxiliary in \eqref{aux-gh} has $d_\phi= 2$.

In general the gauge fixing term 
\be 
\label{GF}
S_{GF} = \frac{1}{2\alpha}\int\!\! \sqrt{\bar{g}}\,\bar{g}^{\mu\nu} F_\mu F_\nu\,,
\ee
breaks background rescaling invariance. However the $\hmn$ transformation \eqref{rescale1} drops out of any legitimate gauge fixing,\footnote{Imposing $F_\mu=0$ should project out gauge transformations only. This is only possible if all terms contain covariant derivatives of $\hmn$.} since $\bar{\nabla}_\alpha \bgmn =0$. Therefore only \eqref{rescalebg} makes a difference. If we furthermore restrict to gauges where $F_\mu$ scales homogeneously under \eqref{rescalebg} (this includes all the usual gauges) then  background rescaling effectively just changes the gauge parameter $\alpha$ in \eqref{GF}.

It will be useful to make the typical choice which is that of De Witt gauge  \cite{Codello:2007bd,Machado:2007,Benedetti:2012,Demmel:2014hla,Demmel:2014fk,Demmel2015b}:
\be 
\label{F}
F_\mu = \bar{\nabla}_\rho h^\rho_\mu - \frac{
1}{d}\bar{\nabla}_\mu h^\rho_\rho\,.
\ee
We see easily that this is indeed invariant under \eqref{rescale1}. Thus using \eqref{rescalebg}, noting the inverse metric hidden in $F_\mu$,  we have that $\delta F_\mu = 2 F_\mu$. Therefore altogether background rescaling has the effect of changing the gauge fixing term \eqref{GF} as
\be 
\label{rescaleGF}
\delta S_{GF} = \left(6-d\right) S_{GF}\,,
\ee
and this in turn can be regarded as a change in the gauge fixing parameter: $\delta \alpha = (d-6)\,\alpha$. Since in the literature, Landau gauge  is chosen by sending $\alpha\to0$ at a point in the calculation when this limit is unambiguous (see \eg \cite{Benedetti:2012,Demmel:2014hla}), it would appear that this actually has no effect, and thus for this gauge, background rescaling invariance is actually respected by the gauge fixing term.
This last statement is actually true, however in treating this limit carefully we will see that we have to alter the transformation laws for the gauge degrees of freedom $\xi_\mu$ and $\sigma$. 

Substituting the York decomposition \eqref{TT}, the physical fields $h^T_{\mu\nu}$ and $h$ drop out of \eqref{F}, leaving only dependence on  $\xi_\mu$ and $\sigma$. Turning our attention to the rest of the action, we note that at the linearised level, a diffeomorphism invariant action does not depend on $\xi_\mu$ and $\sigma$ since they parameterise linearised gauge transformations.
However since they parametrise only the linearised piece of the gauge transformations, beyond the linearised level such an action does depend on $\xi_\mu$ and $\sigma$. Furthermore after gauge fixing, in reality neither the rest of the bare action nor the rest of the effective action is diffeomorphism invariant if this is expressed through $\hmn$, since this is replaced by BRST invariance.\footnote{Once the IR cutoffs are in place even this is broken. Invariance is then expressed through modified Ward identities.} We have seen in sec. \ref{sec:WI} that for the York decomposition to be able to respect background rescaling invariance we require  $\xi_\mu$ and $\sigma$ to be invariant. However in the limit of very small $\alpha$, all dependence on $\xi_\mu$ and $\sigma$  can be neglected in comparison to that coming from \eqref{GF}. Since \eqref{GF} is bilinear in $\xi_\mu$ and $\sigma$, but actually divergent in the limit $\alpha\to0$, it thus follows that to restore background rescaling invariance we must actually choose $\xi_\mu$ and $\sigma$ to transform homogeneously so as to absorb the change \eqref{rescaleGF}.
We thus see that in Landau gauge, $\xi_\mu$ and $\sigma$ transform as $\delta u = (d-6) u/2$, \ie they satisfy \eqref{rescaleu} with $d_{\rm gauge}=d_{\rm ghost} = 6$.

Finally, the physical component fields in \eqref{TT} remain with the transformation laws we already established in sec. \ref{sec:WI}. Thus they
also satisfy \eqref{rescaleu} but with $d_{h^T_{\mu\nu}} =d$ (thus making it invariant), and $d_h = d_{\bar{h}} = d-4$, except that also $\delta h$ and $\delta \bar{h}$ have the inhomogeneous parts in \eqref{rescale2} and \eqref{rescale3}. Unlike the auxiliary and ghost fields, and the gauge degrees of freedom in Landau gauge, the $d_u$ for these component fields are not there to ensure that the bilinear terms are invariant. Rather in this case, the higher order $\bar{h}$ interactions restore invariance. At the exact level this is achieved along the lines discussed at the end of sec. \ref{sec:WI}. At the modified level this is achieved in the way we are about to derive.

\section{IR cutoff terms under background rescaling}
\label{sec:IR}

In the literature \cite{Demmel:2014fk,Demmel2015b,Ohta2016,Falls:2016msz}, the
 IR cutoff $k$ is implemented through replacing the appropriate Laplacian $\Lap$ with
 \be 
 \label{P}
 P_k(\Lap) = \Lap + k^2r(\Lap/k^2)\,.
 \ee
Here $r(z)$ is a dimensionless cutoff profile, which suppresses modes with $z<1$. Since $\Lap$ transforms as \eqref{rescaleDelta}, we can quantify the breaking of background rescaling invariance. Writing RG time as $t=\ln(k/\mu)$ with $\mu$ some fixed physical scale, we have:
\be 
\label{rescaleP}
\delta P_k(\Lap) = 2\, P_k(\Lap) -\partial_t P_k(\Lap)\,.
\ee 
Since $P_k(\Lap)$ is still a map from the space of fluctuations $u$ back into itself,
it follows that its transformation law merely reflects dimensional assignments (as explained in sec.  \ref{sec:symmetries}). Indeed if we allowed $k$ to transform as 
\be 
\label{rescalek} 
\delta k=k\,, 
\ee
only the first term would have appeared. The second term is therefore there in effect to untie this transformation on $k$ and thus leave $k$ invariant.

The actual infrared (IR) cutoff $\cR$, which is added by hand to the bilinear terms, involves further dependence on $k$ and $\bgmn$, and is constructed to implement the replacement \eqref{P} in the Hessian $\hess_u$ for each type of fluctuation $u$. Following standard practice, we have introduced the shorthand 
\be 
\label{hessian}
\Gamma^{(2)}_u:=\frac{1}{\sqrt{\bar{g}}(x) \sqrt{\bar{g}}(y)} \frac{\delta^2\Gamma}{\delta u(x)\,\delta u(y)}\,,
\ee
and treat it as a differential operator in the following.
The bar over the Hessian, as in $\hess_u$, denotes the further standard step that this is evaluated on the background, \ie all fluctuation fields are then set to zero.
In other words, the cutoff is given by
\be 
\label{IRcutoff}
\cR_u = \hess_u(P_k) -\hess_u(\Lap)\,.
\ee
The label $u$ on the cutoff  serves as a reminder that not only the form of the cutoff but also the form of the Hessian, the cutoff profile $r_u$ in \eqref{P}, and the Laplacian $\Lap$, will in general depend on the choice of fluctuation field.
In particular the Hessian, and thus also the IR cutoff, carry indices as appropriate for the given fluctuation field $u$, which we do not display explicitly.
Again by the arguments of sec. \ref{sec:symmetries} and above, we know that $\cR$, like the Hessian $\hess$ itself when evaluated on the background, scales homogeneously together with a correction for the $k$ dependence:
\be 
\label{rescaleIR}
\ie\qquad \delta \hess_u = \left(d_{\cR u}   -\partial_t\right)\! \hess_u\qquad\mathrm{and}\qquad
\delta \cR_u = \left(d_{\cR u}   -\partial_t\right)\! \cR_u\,.
\ee
And again by \eqref{rescaleTensor}, the index $d_{\cR u}$ differs from the dimension if $\hess$ carries indices:
\be 
d_{\cR u} = [\hess_u]+ p_u-q_u\,,
\ee
where $[\hess_u]$ is the mass dimension of the Hessian, and $p_u$ and $q_u$ the number of upper and lower indices respectively. 

For the auxiliary, gauge and ghost fields this works out to be nothing but 
$d_{\cR u} = d_u$, as was in effect arranged above to be the case by requiring background scale invariance. For the graviton $h^T_{\mu\nu}$,
the result is fixed in all cases in the literature by factoring out Newton's constant so that $[\hess]=2$. Taking into account the four upstairs indices (\aka two $\bar{g}^{\mu\nu}$) that are required to contract indices on a pair of $h^T_{\mu\nu}$, we see that this implies that the index $d_{\cR{h^T_{\mu\nu}}}=6$. 
From \eqref{TT} we see that for $\xi_\mu$,  we increase $[\hess]$ by 2 but lose two contravariant indices, and again for $\sigma$ we increase $[\hess]$ by 2 and lose two contravariant indices. Thus the index we would deduce for these gauge degrees of freedom is also $d_{\cR u}=6$. In fact the transformation law for these fields has already been determined in sec. \ref{sec:GF} by requiring invariance of the gauge fixing term in the Landau gauge limit, where however we also found $d_{\cR u}=6$. Finally for $h$, we just lose all contravariant indices, implying $d_{\cR h} =2$. This last index could also have been read off from the $O(\bar{h}^2)$ part of the example \eqref{invariant-example}.


Putting all this together we have thus shown that in Landau gauge, the IR cutoff terms  
\be 
\label{IRcutoffTerms}
S_{\cR} = \frac{1}{2}\int\!\sqrt{\bar{g}}\, \sum_u\! u \cR_u u
\ee
(where the sum is over all fluctuation fields) are the only terms that violate background scaling invariance. We have shown that, for all the implementations in the literature, they transform as
\bea 
\delta S_{\cR} &=& 2d\! \int\! \sqrt{\bar{g}}\, \cR_h h \,+\,  \frac{1}{2}\int\!\sqrt{\bar{g}}\, \sum_u\! u\, \left( d_{\cR u} - d_u -\partial_t \right) \! \cR_u \,u \nonumber\\
&=& 2d\! \int\! \sqrt{\bar{g}}\, \cR_h h \,-\frac{1}{2}\int\!\sqrt{\bar{g}}\, \sum_u\! u\,  \dot{\cR}_u \,u+ \frac{6-d}{2}\!\!\sum_{u=h^T_{\mu\nu}, h}\! \int\!\sqrt{\bar{g}}\,  u \cR_u u \,.
\label{breaking}
\eea
In the first line we have used \eqref{rescalebg}, \eqref{rescale2}, \eqref{rescaleu} and \eqref{rescaleIR}. 
In the second line we have written $\partial_t$ as an over-dot,   and recognised that $d_{\cR u} - d_u$ is non-vanishing only for the physical fields where in both cases we find $d_{\cR u} - d_u=6-d$.

\section{The modified Ward identity}
\label{sec:mWI}


Having established the transformation laws for the fields in sec. \ref{sec:WI} and \ref{sec:GF}, 
and the way that the invariance is broken by IR cutoffs, as displayed in \eqref{breaking}, the derivation of the broken (\aka modified) Ward identity is standard and straightforward. We sketch the steps. 

The partition function \eqref{Z} needs to be replaced by one that includes the York decomposition \eqref{TT}, and the gauge fixing term \eqref{GF}, and thus all the ghost and auxiliary fields that follow from this -- in whatever implementation the reader prefers. Again using $u$ to denote all the fluctuation fields, the source term in \eqref{Z} now appears schematically as $ \sum_u\int\! J_u u$. Rescaling the background metric by \eqref{rescalebg}, and compensating through transforming the fluctuation fields via \eqref{rescale2} and \eqref{rescaleu}, we see that  background rescaling is equivalent to transformation of the source terms and IR cutoff terms. Thus:
\begin{multline} 
-2 \int\! \bgmn\, \frac{\delta W}{\delta \bgmn} = 2 d \!\int\! J_h +\frac{1}{2}\sum_u (d-d_u)\int J_u \frac{\delta W}{\delta J_u} \\
-2d\!\int\!\sqrt{\bar{g}}\, \cR_h \frac{\delta W}{\delta J_h}
 \,+\, \frac{1}{2}\int\!\sqrt{\bar{g}}\, \sum_u\! \frac{\delta W}{\delta J_u}\dot{\cR}_u \frac{\delta W}{\delta J_u}
 \,+\, \frac{d-6}{2}\!\!\sum_{u=h^T_{\mu\nu}, h}\! \int\!\sqrt{\bar{g}}\,  \frac{\delta W}{\delta J_u} \cR_u \frac{\delta W}{\delta J_u}\\
  \,+\, \frac{1}{2}\sum_u \mathrm{tr}\! \left(\! \dot{\cR}_u\, \frac{\delta^2 W}{\delta J_u \delta J_u} \right)
 \,+\, \frac{d-6}{2}\!\!\sum_{u=h^T_{\mu\nu}, h}\!\! \!\mathrm{tr}\! \left(\!\cR_u\, \frac{\delta^2 W}{\delta J_u \delta J_u} \right) \,.
 \label{WmWI}
\end{multline}
In the last line we represent the space-time trace of the product of two kernels in effectively the standard way through the DeWitt shorthand. Now we transform to the Legendre effective action 
\be
\Gamma^\mathrm{tot} = -W + \sum_u\int\! J_u u\,, 
\ee
where now $u= \delta W/\delta J_u$ is the classical field, and split off the cutoff terms:
\be 
\Gamma^\mathrm{tot}  = \Gamma + \frac{1}{2}\int\!\sqrt{\bar{g}}\, \sum_u\! u \cR_u u\,.
\ee
Under the transformation \eqref{breaking}, the latter term reproduces the middle line of \eqref{WmWI}, and thus we derive: 
\begin{multline}
\label{mWI}
2\!\int\! \bgmn\,\frac{\delta\Gamma}{\delta\bgmn} \,-\, 2d \!\int\frac{\delta\Gamma}{\delta h}\,-\,
\frac{1}{2}\sum_u (d-d_u)\!\int\! u\, \frac{\delta \Gamma}{\delta u} \\
= \frac12\sum_u\! \mathrm{tr}\!\left[ (\Gamma^{(2)}_u+\cR_u)^{-1}\, \dot{\cR}_u\right] 
\,+\,\frac{d-6}{2}\!\!\sum_{u=h^T_{\mu\nu}, h}\!\! \!\mathrm{tr}\!\left[ (\Gamma^{(2)}_u+\cR_u)^{-1}\, \cR_u\right] \,,
\end{multline}
where the Hessian (with non-vanishing fluctuation fields) is defined in \eqref{hessian}.
Notice that since $d_h = d-4$, the LHS (left hand side) of \eqref{mWI} contains the unbroken Ward identity \eqref{hWI}. Therefore \eqref{mWI} is indeed the hoped-for mWI, namely the Ward identity modified by the addition of terms involved in the gauge fixing and regularisation.

\section{Compatibility of the mWI after approximation}
\label{sec:compatibility}

In this section we address the extent to which the mWI \eqref{mWI} is compatible with the exact RG flow equation \cite{Nicoll1977,Wetterich:1992,Morris:1993} which in this context \cite{Reuter:1996} and this notation, takes the form:
\be 
\label{flow}
\dot{\Gamma} = \frac12\sum_u\! \mathrm{tr}\!\left[ (\Gamma^{(2)}_u+\cR_u)^{-1}\, \dot{\cR}_u\right] \,,
\ee
where again we use \eqref{hessian}.
By compatibility we mean the following \cite{Labus:2016lkh}. Write the mWI in the form $\mathcal{W}=0$ and assume that this holds at some scale $k$. Computing $\dot{\mathcal{W}}$ by using the flow equation, we say that the mWI is compatible if $\dot{\mathcal{W}}=0$ then follows at scale $k$ without further constraints.  

Since the mWI and the exact RG flow equation are both derivable consequences from the path integral representation of the partition function augmented by adding to the action the generic IR cutoff terms \eqref{IRcutoffTerms},
they are formally\footnote{\ie to the extent that this functional integral actually makes sense without further modification/regularisation} guaranteed to be compatible.

On the other hand to make progress we need to approximate the mWI and flow equation. Then their compatibility is far from guaranteed. For example it was shown in ref. \cite{Labus:2016lkh} in the context of
conformally truncated gravity, that after approximating by nothing more than a derivative expansion, background independence as expressed through the modified shift Ward identity, is compatible with the flow equation if \emph{and only if} a power-law cutoff profile is used. Although \textit{a priori} an incompatible system of mWI and flow equation could still have solutions \cite{Labus:2016lkh},  we also showed that in practice there are no consistent solutions to the system in this case. Therefore before we can make further progress and analyse the consequences of imposing \eqref{mWI}, we need to show that it can be compatible with \eqref{flow} within some suitable approximation. 

As seen in ref. \cite{Labus:2016lkh}, the exact compatibility relies on the symmetry of some two loop diagrams, where one of the loops contains the kernel $\dot{\cR}$ from the RHS of the flow equation and the other loop contains the kernel from the RHS of the modified Ward identity and thus also involves the undifferentiated $\cR$. The problem is that the symmetry of the two loop diagram is generically broken by uncontrolled approximations,\footnote{By uncontrolled approximation we mean one where $O(1)$ terms are neglected. 
By contrast a Taylor expansion in a small quantity, \eg a coupling, is a controlled approximation. The symmetry would then be preserved order by order in this small quantity.} including the derivative expansion itself. The symmetry was recovered there by choosing power law cutoff profile because as a consequence of $\dot{\cR}\propto \cR$, the kernels 
themselves then become proportional as functions of the implied internal momenta (that is the internal momentum that is integrated over in forming the space-time trace and in the two loop diagrams). 

We can therefore anticipate that, after approximation, compatibility here will also require that the RHSs of \eqref{mWI} and \eqref{flow} become proportional.\footnote{Indeed  since only bilinear dependence on the ghosts and auxiliary fields is kept (as recalled in the next section), they propagate only in one of the two loops. This already breaks the symmetry.}
 If $d\ne6$, we see already that we will have a problem. If, as in ref. \cite{Labus:2016lkh}, we  try to tackle this by choosing a power law cutoff profile $r_u$ so that $\dot{r}_u\propto r_u$ for the physical fields ($u=h^T_{\mu\nu}, h$), we see this cannot solve the problem since here the infrared cutoff  \eqref{IRcutoff} also depends on $t$ through the terms in the Hessian. 
This rules out a solution using power-law cutoff profile even for the scheme used in ref. \cite{Demmel:2014hla} where only the physical fields themselves make a contribution to the flow equation. For other schemes  \cite{Benedetti:2012,Ohta2016,Falls:2016msz} it 
furthermore cannot make the full RHSs proportional because the contribution from
all the other fluctuation fields (auxiliary, gauge and ghosts) already appear in exactly equal ways on the RHSs of both \eqref{mWI} and \eqref{flow}. Thus  we see that the only way to make these RHSs proportional, and thus compatible in an uncontrolled approximation \cite{Labus:2016lkh},
is to choose space-time dimension $d=6$. Remarkably however, since $d=6$ is already sufficient to make the RHSs identical, compatibility turns out to be guaranteed whatever cutoff profile $r_u$ is used and almost whatever further approximation we impose in computing the right hand sides!

\section{Single metric approximation}
\label{sec:single-metric}

Let us now use the standard  approximations \cite{Reuter:1996}. Then in sec. \ref{sec:compatibility-singlemetric} we will prove that indeed
the mWI remain compatible with the flow equation if and only if $d=6$.
We must first define what we mean by these approximations  in this context. As usual we will take the ghost and auxiliary effective actions to be given by their bare ones (\ie ansatz that they do not flow) and after forming the Hessians, discard dependence on these fields. This therefore just defines the contributions from these fields to the RHS of \eqref{flow}, and similarly now also the RHS of \eqref{mWI}. It also means that the corresponding $u\, \delta\Gamma/\delta u$ terms now vanish on the LHS of \eqref{mWI}. In the literature, for metric fluctuations the single-metric approximation is made, which amounts to replacing
\be 
\label{single-metric}
\frac{\delta^2 \Gamma}{\delta \hmn \delta h_{\alpha\beta}} \quad\mapsto\quad \frac{\delta^2 \Gamma}{\delta \bgmn \delta \bar{g}_{\alpha\beta}}
\ee
on the RHS of \eqref{flow}, followed by discarding all dependence on $\hmn$ (or its York decomposition). This is the one place where we will be slightly less drastic. We will retain dependence on one small part of the physical fluctuation field only, namely $\bar{h}$, the constant part of $h$. Since we still make the single-metric step \eqref{single-metric}, this dependence does not alter the form of the RHSs of \eqref{mWI} and \eqref{flow}. On the LHS of \eqref{mWI} it means we discard all fluctuation field derivatives except those relating to $\bar{h}$. To derive what remains, note that from \eqref{h-expand}  we have
\be 
\label{barh-relation}
\frac{\partial\Gamma}{\partial\bar{h}} = \int\!\!\frac{\delta \Gamma}{\delta h}\,.
\ee
We can also invert the relation \eqref{h-expand} to get:
\be 
\label{invert-h-expand}
\bar{h} = \frac{1}{V}\int\!\!\sqrt{\bar{g}}\,h\,,\quad h^\perp = h - \frac{1}{V}\int\!\!\sqrt{\bar{g}}\,h\,,\quad\mathrm{where}\quad V := \int\!\!\sqrt{\bar{g}}\,.
\ee
Since
\be 
\frac{\delta \Gamma}{\delta h(x)} = \int_y \frac{\delta h^\perp(y)}{\delta h(x)}\frac{\delta\Gamma}{\delta h^\perp(y)}+\frac{\delta\bar{h}}{\delta h(x)}\frac{\partial\Gamma}{\partial\bar{h}}\,,
\ee
(where $\int_y \equiv \int\!d^dy$) we thus find 
\be 
\frac{\delta \Gamma}{\delta h(x)}  = \frac{\sqrt{\bar{g}}}{V}\frac{\partial\Gamma}{\partial\bar{h}} +\frac{\delta \Gamma}{\delta h^\perp(x)} -\frac{\sqrt{\bar{g}}}{V}\int_y\frac{\delta \Gamma}{\delta h^\perp(y)}\,.
\ee
Thus we confirm \eqref{barh-relation} and also find
\be 
 \int\!\! h\frac{\delta \Gamma}{\delta h} = \bar{h}\frac{\partial\Gamma}{\partial\bar{h}}+\int\!\! h^\perp\frac{\delta \Gamma}{\delta h^\perp}\,.
\ee
Since we discard dependence on $h^\perp$, we find finally that the LHS of the mWI \eqref{mWI}
collapses to:
\be 
\label{LHS}
2\!\int\!\! \bgmn\,\frac{\delta\Gamma}{\delta\bgmn} \,-\, 2d \frac{\partial\Gamma}{\partial\bar{h}} -2 \bar{h}\frac{\partial\Gamma}{\partial\bar{h}} = \cdots \,,
\ee
where $\Gamma$ is now only a functional of $\bgmn$ and a function of $\bar{h}$ and $k$.
This is exactly what we would expect to find, given that (apart from an overall sign) it generates the rescaling transformations \eqref{rescalebg} and \eqref{rescale3}.

\section{Compatibility in single metric approximation}
\label{sec:compatibility-singlemetric}

In preparation for the proof of compatibility let us call the linear operator  that generates background rescaling, $\omega$, so that the LHS of \eqref{LHS} is merely $-\omega \Gamma$. Let us write the RHS compactly also so that, on taking the LHS over to the RHS, the mWI as a whole can be written:
\be 
\label{mWI-compact}
0 = \mathcal{W} := \omega \Gamma +  \half\mathrm{tr}[ \triangle \cK ]\,. \ee
Here we write the kernel for the mWI as:
\be 
\label{K}
\cK_u = \dot{\cR}_u + (d-6)\cR_u \delta_{u=\physical}\,,
\ee
recognising that the correction in \eqref{mWI} is non-vanishing only for physical fields $h$ and $h^T_{\mu\nu}$. We write the full propagator as
\be 
\triangle_u := (\Gamma^{(2)}_u+\cR_u)^{-1}
\ee
(where the use of the triangle symbol, $\triangle$, is not to be confused with $\Delta$ as in the background Laplacian $\Lap$).  Similarly we write the flow equation \eqref{flow} more compactly as
\be 
\label{flow-compact}
\dot{\Gamma}=\half \mathrm{tr}[ \triangle \dot{\cR}]\,.
\ee
Finally the reader should understand that all terms in the space-time trace carry a $u$ label which is summed over. We drop this label because the cancellations we are about to see actually happen for each species separately, so this extra structure will play no r\^ole.

In fact we will shortly be interested in the further approximation that comes about from choosing a compact maximally symmetric background space (the Euclidean four-sphere). In this case the expressions can be further simplified by summing over the eigenmodes of the appropriate Laplacians. However the domain of compatibility of the mWI with the flow equation is unchanged by this specialisation so we will furnish the proof for a general background metric. Note also that we make no assumption on the form of the cutoff profile $r_u$ in the following. Again the domain of compatibility is unchanged by this choice. In practice, since we want to adopt the flow equations derived in the literature we will typically be interested in the optimised cutoff profile \cite{opt1,opt3}.

Finally, taking the RG time derivative of the mWI \eqref{mWI-compact} and substituting the flow equation \eqref{flow-compact}, we get
\be 
\dot{\mathcal{W}}= \half \omega \mathrm{tr}[ \triangle \dot{\cR}]+
\half \mathrm{tr}[ \triangle \dot{\cK}]-\half \mathrm{tr}[ \triangle (\dot{\Gamma}^{(2)}+\dot{\cR})\triangle\cK]\,.
\ee
We can evaluate the action of the linear operator $\omega$ on the first term, since by \eqref{rescaleIR} we have $\omega \dot{\cR} = (d_{\cR} -\partial_t)\dot{\cR}$ and 
\be 
\omega \triangle = -\triangle( \omega\Gamma^{(2)}+\omega\cR)\triangle =-d_{\cR}\triangle+\triangle\left([d_{\cR}-\omega]\Gamma^{(2)}+\dot{\cR}\right)\triangle\,.
\ee
Note that from \eqref{rescaleIR} we have that $[d_{\cR}-\omega]\Gamma^{(2)}= \dot{\Gamma}^{(2)}$ whenever the Hessian contains only the background metric, \ie $\Gamma^{(2)}=\hess$. However this latter equality is in general not true for the physical fluctuations since there we retain dependence on $\bar{h}$.
Collecting terms we thus have that
\be
\dot{\mathcal{W}}= \frac12 \mathrm{tr}\!\left[ \triangle (\dot{\cK}-\ddot{\cR})\right]
-\frac12 \mathrm{tr}\!\left[ \triangle \dot{\cR}\triangle\left(\cK-\dot{\cR}\right)\right] 
+\frac12 \mathrm{tr}\!\left[ \triangle\left( [d_{\cR}-\omega]\Gamma^{(2)}\right)\triangle\dot{\cR}\right]
-\frac12 \mathrm{tr}\!\left[ \triangle \dot{\Gamma}^{(2)}\triangle\cK\right]\,.
\ee
According to our definition of compatibility (\cf sec. \ref{sec:compatibility}) this must evaluate to zero without further conditions. By substituting for $\omega \Gamma^{(2)}$ and $ \dot{\Gamma}^{(2)}$, using the mWI and flow equation respectively, we would be led to the approximated two-loop diagrams discussed in sec. \ref{sec:compatibility}, which we have already seen will fail to cancel unless $\cK = \dot{\cR}$. On the other hand, we see from above that if $\cK = \dot{\cR}$, then $\dot{\mathcal{W}}=0$ will follow, provided that it can be shown that $[d_{\cR}-\omega]\Gamma^{(2)}= \dot{\Gamma}^{(2)}$.  In sec. \ref{sec:background-independence} we will see that 
this is indeed a consequence, since the equations imply that the solution satisfies $\Gamma^{(2)}=\hess$.
  We will have thus shown that the mWI and the flow equation are compatible if and only if $\cK = \dot{\cR}$. By \eqref{K} this means they are compatible if and only if we work in $d=6$ space-time dimensions.

\section{The significance of six}
\label{sec:six}

In this section we pause for a moment to give an intuitive explanation for the need to impose $d=6$ dimensions from here on.
Although, as discussed in sec. \ref{sec:compatibility}, the exact flow equation and exact mWI for background rescaling invariance, are automatically compatible, we emphasise that this is typically no longer true when we make uncontrolled approximations.
In fact we have seen that once such approximations are made, the mWI and flow equation will be compatible if and only if we choose $d=6$ spacetime dimensions.

The price we pay for working within an uncontrolled approximation scheme is that we must set $d=6$. 
If we do not maintain compatibility within the approximation scheme itself then, as discussed in sec. \ref{sec:compatibility} and in ref. \cite{Labus:2016lkh}, we would find no solutions at all to the combined system of mWI and flow equations. 

In sec. \ref{sec:GF} we saw that, after suitable choices of background scaling dimension $d_u$, all fluctuation fields, apart from the physical fields, have Hessians whose actions are invariant under background rescaling. (This is true for the gauge degrees of freedom $\xi_\mu$ and $\sigma$ only after the Landau gauge limit is taken.) In contrast the action for the Hessian for physical fluctuations transforms with no homogeneous part only in $d=6$ dimensions, as we will prove in the next section.
This is why the difference between the RHSs of the mWI \eqref{mWI} and flow equation \eqref{flow} disappears in precisely $d=6$ dimensions. 
To understand intuitively why six dimensions is singled out,  consider a term of the form
\be 
\frac{1}{2}\int\!\sqrt{\bar{g}}\, \bar{g}^{\alpha\mu}\bar{g}^{\beta\nu} h^T_{\alpha\beta} (-\bar{\nabla}^2) h^T_{\mu\nu}\,,
\ee
where of course $\bar{\nabla}^2 = \bar{g}^{\sigma\rho} \bar{\nabla}_\sigma \bar{\nabla}_\rho$. Counting powers of the background metric we see indeed that invariance under \eqref{rescalebg-1} requires exactly  $d=6$ dimensions. ($h^T_{\mu\nu}$ itself is invariant as established below \eqref{rescale2}.) The analysis in secs. \ref{sec:symmetries}, \ref{sec:GF} and \ref{sec:IR} then establishes there is no homogeneous part whatever the form of the Hessian, provided only that Newton's constant is factored out, as is always done in the literature. Factoring out Newton's constant ensures that the dimension of the Hessian is $[\hess]=2$ and results in a theory based on second order derivative terms as illustrated above. In contrast for example, had we based the theory on a four-derivative action such as in Weyl gravity, we would find that $d=8$ is singled out instead.

\section{Simultaneous solution and scale independent variables}
\label{sec:background-independence}

Since we need compatibility to make further progress (\cf the discussion in sec. \ref{sec:compatibility} and the last section), from now on we specialise to the case of $d=6$ dimensions. It is remarkable that compatibility is regained in this case, and even more remarkable that this is so for almost any approximation. 

Indeed although we wish to apply these results to the standard procedures and approximations in the literature, in particular for the optimised cutoff  \cite{opt1,opt3} and forming $f(R)$ approximations by projecting on a maximally symmetric background metric, we saw in sec. \ref{sec:compatibility-singlemetric}, that compatibility will hold in $d=6$ dimensions whatever cutoff profiles $r_u$ we choose, and whatever background metric we choose, provided only that the background space-time is compact (has finite volume) so that $\bar{h}$ is well defined, \eg through \eqref{invert-h-expand}. We will now see that the solution of the mWI in terms of new scale independent variables is sufficiently powerful that it also holds independent of the choice of cutoff profile and independent of the choice of $\bgmn$.

We set up in sec. \ref{sec:single-metric} a slightly extended version of the single-metric approximation, in that we keep also dependence on $\bar{h}$. Having shown that the mWI remains compatible with the flow equation in this case we now show that under these circumstances we can solve  these two equations simultaneously to derive background scale independent variables. These steps are inspired by the discovery of background independent variables in previous cases \cite{Bridle:2013sra,Dietz:2015owa,Labus:2016lkh} but due to the much weaker nature of the mWI we impose, we obtain not background independence here but only independence from the overall scale of the background metric.

Again the key is to combine the mWI and flow equation by eliminating the non-linear pieces on the RHSs, after which the linear equation may be solved by the method of characteristics. Combining the flow and mWI we have, by \eqref{LHS}, the linear partial differential equation:
\be 
\label{linearPDE}
\dot{\Gamma} \,+\, 2d \frac{\partial\Gamma}{\partial\bar{h}} +2 \bar{h}\frac{\partial\Gamma}{\partial\bar{h}} \,-\, 2\!\int\!\! \bgmn\,\frac{\delta\Gamma}{\delta\bgmn} =0\,.
\ee
The first term implies that its characteristic curves can be parametrised by the RG time $t$ itself.\footnote{Had we parametrised with $s$ say, then the first term would imply $dt/ds=1$.} After this the vector field generating the characteristic curves is just the one we derived in the unbroken case, namely \eqref{charGamma}, \eqref{charbg} and \eqref{charhbar}. Thus the solution is again that $\Gamma$ is constant for the characteristics defined by \eqref{solbg} and \eqref{solhbar}. However, now that the characteristic curve auxiliary parameter is endowed with extra meaning, being identified with the RG time, we need to interpret them differently. \emph{Indeed the left hand side of \eqref{linearPDE} is $\hat{\omega}\Gamma$, where the extended background rescaling operator $\hat{\omega} = \omega + \partial_t$
just generates background rescaling transformations such that $k$ now also participates as in \eqref{rescalek}.}  Following the appendix to ref. \cite{Dietz:2015owa}, and also following ref. \cite{Labus:2016lkh}, we thus rewrite the integration constant for $\bar{h}(t)$ in \eqref{solhbar} as an integration constant 
\be 
\label{that}
\hat{t}= t -\half \ln(1+\bar{h}/d)
\ee
for $t$ (where strictly now $d=6$). This then defines the background rescaling invariant version of cutoff scale $\hat{k}$ as: 
\be 
\label{khat}
\hat{k} = k/\sqrt{1+\bar{h}/d}\,.
\ee
Remembering that $k$ now also transforms, and using 
\eqref{rescale3}, we see that indeed this new form of cutoff scale is invariant under background rescaling.
From \eqref{solbg} or directly from \eqref{invariant}, we can define the background rescaling invariant version of the background metric as:
\be 
\label{ghat}
\hat{g}_{\mu\nu}(x) = \left(1+\frac{\bar{h}}{d}\right) \bgmn(x)\,.
\ee
Then the solution $\Gamma \equiv \Gamma_k[\bgmn](\bar{h})$ to \eqref{linearPDE} can be written in terms of a new functional $\hat{\Gamma}$ that is invariant along the characteristics:
\be
\label{Ghat}
\Gamma 
= \hat{\Gamma}_{\hat{k}}[\hat{g}_{\mu\nu}]\,,
\ee
in which all explicit reference to $\bar{h}$ has disappeared. By differentiating this with respect to $\bar{h}$, and using the definitions \eqref{that} and \eqref{ghat}, it is straightforward to verify that this does indeed solve \eqref{linearPDE}. Since their LHSs now agree and their RHSs are anyway equal (in $d=6$), this immediately implies that the flow equation \eqref{flow} and mWI \eqref{mWI}  reduce to the same equation for $\hat{\Gamma}$. 

To find this equation, we note that the LHS follows from the equality:
\be 
\label{t-deriv-equality}
\partial_t|_{\bar{h},\bar{g}_{\mu\nu}} \Gamma = \partial_{\hat{t}}|_{\hat{g}_{\mu\nu}} \hat{\Gamma}\,.
\ee
On the RHS we use the fact that the change of variables \eqref{khat} and \eqref{ghat} is in the form of a ($\bar{h}$-dependent) finite background rescaling transformation where now $k$ actively participates: the corresponding infinitesimal transformations being \eqref{rescalebg} and \eqref{rescalek}. 

Explicitly, consider any quantity $\bar{Q} :=Q(\bgmn,k)$ that under these infinitesimal transformations,  transforms homogeneously as $\delta \bar{Q} = d_Q\, \bar{Q}$. Under the change of variables \eqref{khat} and \eqref{ghat}, $\bar{Q}$ thus becomes
\be 
\bar{Q} =  \hat{Q} \left(1+\frac{\bar{h}}{d}\right)^{d_Q/2}\,,
\ee
where by $\hat{Q}$ we mean simply\footnote{We will see at the end of this section how this definition of a hatted quantity is actually consistent with $\hat{\Gamma}$ as already defined in \eqref{Ghat}.} 
\be 
\label{hatQ}
\hat{Q} := Q(\hat{g}_{\mu\nu},\hat{k})\,. 
\ee
Examples are of course $k$ and $\bgmn$ themselves. 
Recalling the explanation below \eqref{rescaleP}, we also see that $P_k(\Lap)$ now scales homogeneously. From \eqref{P} we thus find
\be 
P_k(\Lap) =  P_{\hat{k}}(\hat{\Delta})\left(1+\frac{\bar{h}}{d}\right)\,.
\ee
Carrying this through to the IR cutoffs $\cR$ and Hessians evaluated on the background, we see that they too now transform homogeneously:
\be 
\label{homogeneous-to-hat}
\hess_u = \hat{\Gamma}^{(2)}_u\left(1+\frac{\bar{h}}{d}\right)^{d_{\cR u}/2}\qquad\mathrm{and}\qquad
\cR_u = {\hR}_u \left(1+\frac{\bar{h}}{d}\right)^{d_{\cR u}/2} \,.
\ee
However on the RHS of the flow equation/mWI, these $\bar{h}$-dependent powers just cancel between  $\dot{\hR}$ and the inverse of $[\hat{\Gamma}^{(2)}+\hR]$.

This deals with all the contributions from all the auxiliary fluctuation fields and ghost fields since their Hessians are automatically evaluated on the background, \ie have no other field dependence, and also with the gauge degrees of freedom since in the Landau gauge limit their Hessians also have no other field dependence. 

This leaves the Hessians $\Gamma^{(2)}_u$ for the physical degrees of freedom $u=h^T_{\mu\nu}, h$, since \textit{a priori} \eqref{single-metric} still depends on both $\bgmn$ and $\bar{h}$. Recalling that the Hessians are being regarded as differential operators it is helpful to think of them as embedded in an action
\be 
\label{physical-hessian-action}
\frac{1}{2}\int\!\sqrt{\bar{g}}\, \sum_{u=\physical}\!\! u\, \Gamma^{(2)}_u u\,,
\ee
in order to understand their  transformation properties under background rescaling.

If we replace $\Gamma^{(2)}_u$ by $\hess_u$, \ie set $\bar{h}=0$:
\be 
\label{background-physical-hessian-action}
\frac{1}{2}\int\!\sqrt{\bar{g}}\, \sum_{u=\physical}\!\! u\, \hess_u u\,,
\ee
 then this transforms into a $\bar{h}$ one-point vertex from the transformation \eqref{rescale2} on the explicit $h$s, or equivalently from the transformation \eqref{rescale3} on the explicit $\bar{h}$s. However we now prove that
 the homogeneous part of the transformation has index $6-d=0$, \ie vanishes. 
We have basically already demonstrated this below \eqref{breaking}. Recall that in sec. \ref{sec:WI}, see also sec. \ref{sec:GF}, we established that the graviton $h^T_{\mu\nu}$ is invariant, while $h$ transforms with index $2$; then in sec. \ref{sec:IR} we established that $\hess_u$  transforms homogeneously,  with index $d_{\cR u} =6$ and $2$ respectively. The $t$-derivative correction shown in \eqref{rescaleIR} is once again no longer required since $k$ now actively participates. Taking into account that from \eqref{rescalebg}, $\sqrt{\bar{g}}$ transforms with index $-d$, and adding up all the contributions, we confirm that overall the action \eqref{background-physical-hessian-action}  transforms with index $6-d=0$, \ie with no homogeneous part.

Next we prove that the interaction term involving one extra $\bar{h}$ must vanish. The reasoning basically follows that surrounding eqn. \eqref{O(h-bar)}. Indeed, if the interaction term did not vanish, since the $\bar{h}$ transformation in \eqref{rescale3} is inhomogeneous, it would yield a bilinear term which has nothing to cancel, because we have already shown that the bilinear terms \eqref{background-physical-hessian-action} transform with no homogeneous part. 

Having proved there is no $\bar{h}uu$ interaction, we can similarly show that there is no $\bar{h}^2 uu$ interaction, for if there was, under \eqref{rescale3} it would transform into a $\bar{h}uu$ piece. This piece can only cancel the homogeneous part of a transformed $\bar{h}uu$  interaction. Since the latter does not exist we conclude there is no $\bar{h}^2 uu$ interaction either. Proceeding iteratively, we have thus proved that in $d=6$ dimensions and as a consequence of 
the combined mWI and flow equation, \viz \eqref{linearPDE},  the Hessian for physical fluctuations actually has no $\bar{h}$ dependence, and thus in this case $\Gamma^{(2)}_u=\hess_u$ even for the physical fluctuations.

It immediately follows then that in $d=6$ dimensions, the Hessians for all fluctuations transform as in \eqref{homogeneous-to-hat} and thus the one remaining equation for $\hat{\Gamma}$, which we may legitimately identify as the background scale independent flow equation, simply reads:
\be 
\label{flowhat}
\partial_{\hat{t}}\hat{\Gamma} = \frac12\sum_u\! \mathrm{tr}\!\left[ (\hat{\Gamma}^{(2)}_u+{\hR}_u)^{-1}\, \partial_{\hat{t}}{\hR}_u\right] \,.
\ee
We remind the reader that here the Hessians themselves are defined by the standard suite of approximations \cite{Reuter:1996}. In particular there is no longer the distinction set out in sec. \ref{sec:single-metric} for the physical fluctuations: their Hessians too are evaluated on the background. All that is further required to construct the above equation is simply to replace $\bgmn$ by $\hat{g}_{\mu\nu}$ and $k$ by $\hat{k}$ on the RHS. 
On the LHS, we can see this as taking the $\bar{h}$-dependent $\Gamma \equiv \Gamma_k[\bgmn](\bar{h})$ and using the equality \eqref{t-deriv-equality}. However, since the RHS is now identical to the standard (single-metric) approximations,
we see that actually the solution is then guaranteed to be the same as a single-metric approximation solution $\Gamma_k[\bgmn]$, with $\bgmn$  replaced by $\hat{g}_{\mu\nu}$ and $k$ replaced by $\hat{k}$. In other words, we have proved from the flow equation that the change to background scale independent variables just amounts to replacing $\bgmn$  with $\hat{g}_{\mu\nu}$ and $k$ with $\hat{k}$. 

From here, finally,  we can further approximate by retaining only certain operators, for example all powers of $\hat{R}$, as subsumed in the function $f_{\hat{k}}(\hat{\sR})$. Of course the solutions are therefore identical to those $f_k(\bar{\sR})$ that would have been obtained originally. All that has happened is that $\bgmn$  is replaced by $\hat{g}_{\mu\nu}$ and $k$ replaced by $\hat{k}$.


\section{Conclusions}
\label{sec:conclusions}

From the previous section, we therefore see that we arrive at \emph{precisely the same flow equation as in the single metric approximation}, except that now the background metric $\bgmn$ and the cutoff scale $k$ 
are replaced by $\hat{g}_{\mu\nu}$ and $\hat{k}$ respectively. 
Nevertheless the interpretation of these quantities is crucially different. 

Since $\hat{k}$ depends, through \eqref{khat}, on part of the dynamical field $\hmn$,  the value of $\hat{k}$ depends on the physical situation. The fact that $\bar{h}$ depends on the physical situation is the analogue for the expectation value, of the fact that in the partition function, $\bar{h}$ must integrated over all acceptable values. 
Similarly, the background metric $\bgmn$ is replaced by $\hat{g}_{\mu\nu}$ which, through dependence on $\bar{h}$, is now actually a dynamical quantity. We therefore cannot think of $\hat{g}_{\mu\nu}$ as fixed  but rather must solve with an ensemble of values in mind.

Background rescaling invariance is now built in. From \eqref{ghat} and \eqref{khat},
exactly the same solution refers in fact to an infinite ensemble of background space-times related by an arbitrary finite rescaling $\bgmn\mapsto  \bgmn/\alpha^2$, since this can be compensated by $\bar{h}\mapsto -d +(\bar{h}+d)\alpha^2$ and  $k\mapsto k \alpha$, thus leaving $\hat{g}_{\mu\nu}$ and $\hat{k}$ alone. Since this map changes $\bar{h}$, we also see explicitly how the overall scales of the full metric and background metric get identified with each other after solving for the mWI. 

At first sight, we still have an option to treat preferentially the region where there are no eigenvalues left to integrate out, namely to do so using $\hat{k}<a\sqrt{\hat{R}}$, \ie use the modified Laplacian on the manifold built with $\hat{g}_{\mu\nu}$. Ranging over $\alpha$ as defined above, this corresponds to setting the infrared cutoff $k$ to be different on manifolds of different background curvature.
Indeed in the picture of fig. \ref{fig:allR}, it corresponds to choosing a ray $k = \kappa \sqrt{\bar{R}}$ for some proportionality constant $\kappa$, instead of a horizontal line. The problem is that now when $\kappa$ is lowered, corresponding to  integrating out, the picture changes irrevocably. In particular if $\kappa>a$ before and $\kappa<a$ after, then again we go from a situation where there were eigenvalues to integrate out to one where there are none. Once again no rescaling can return it to its original form. Therefore we cannot arrange the IR cutoff $k$ in this fashion without again destroying the Wilsonian RG.

Instead we realise mathematically the Wilsonian RG picture we set out in fig. \ref{fig:allR}. As we have just reviewed, the basic condition for the Wilsonian RG to make sense is that $k$ must be treated as independent of the background metric $\bgmn$. We cannot impose conditions on the solution that depend on comparing $k$ to a particular choice of $\bgmn$ without violating a precondition for the Wilsonian RG which is that after lowering $k$ and rescaling back to the original size, the same ensemble of space-times plus fluctuations can be recovered. Treating $k$ as independent however, means that under background rescaling $\hat{k}$ is now active: $\hat{k}\mapsto \hat{k}/\alpha$. We thus recover in the $f(\hat{\sR})$ type approximations we have now formulated, the fact that
solutions must remain smooth over the full range of $\hat{k}$, or equivalently over all $\hat{\sR} = \hat{R}/\hat{k}^2$ no matter how large. 

\section*{Acknowledgments}
TRM acknowledges support from STFC through Consolidated Grant ST/L000296/1.




\bibliographystyle{hunsrt}
\bibliography{references} 

\end{document}